\documentclass{article}
\usepackage{graphicx} 
\begin{document}

\title {
Content-Based Sub-Image Retrieval with Relevance Feedback
}

\author{Jie Luo and Mario A. Nascimento$^\star$\\
Dept. of Computing Science, University of Alberta, Canada\\
$^\star$mn@cs.ualberta.ca}

\date{}

\maketitle

\begin {abstract}
The typical content-based image retrieval problem is to find images
within a database that are similar to a given query image.
This paper presents a solution to a different problem, namely that
of content based {\em
sub}-image retrieval, i.e., finding images from a database 
that {\em contains} another image.  Note that this is different from
finding a region in a (segmented) image that is similar to another image
region given as a query. 
We present a technique for CBsIR that explores relevance feedback, i.e., 
the user's input on intermediary results,
in order to improve retrieval efficiency.
Upon modeling images as a set of overlapping and recursive tiles, 
we use a tile 
re-weighting scheme that assigns penalties to each tile
of the database images and updates the tile penalties for all relevant
images retrieved at each iteration
using both the relevant and irrelevant images identified by the user.
Each tile is modelled by means of its color content using a compact
but very efficient method which can, indirectly, capture some notion
of texture as well, despite the fact that only color information is
maintained.
Performance evaluation on a largely heterogeneous dataset of 
over 10,000 images shows that the system can achieve a stable average 
recall value of 70\% within the top
20 retrieved (and presented) images
after only 5 iterations, with each such iteration taking about
2 seconds on an off-the-shelf desktop computer.
\end {abstract}

\section {Introduction}
\begin{figure*}
\center
\begin{tabular}{c | c}
Query Image & Answer Set \\
\includegraphics[width=0.8cm]{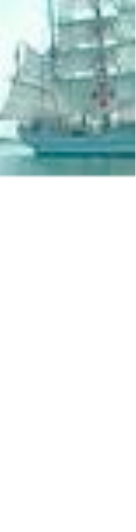} &
\includegraphics[width=10cm]{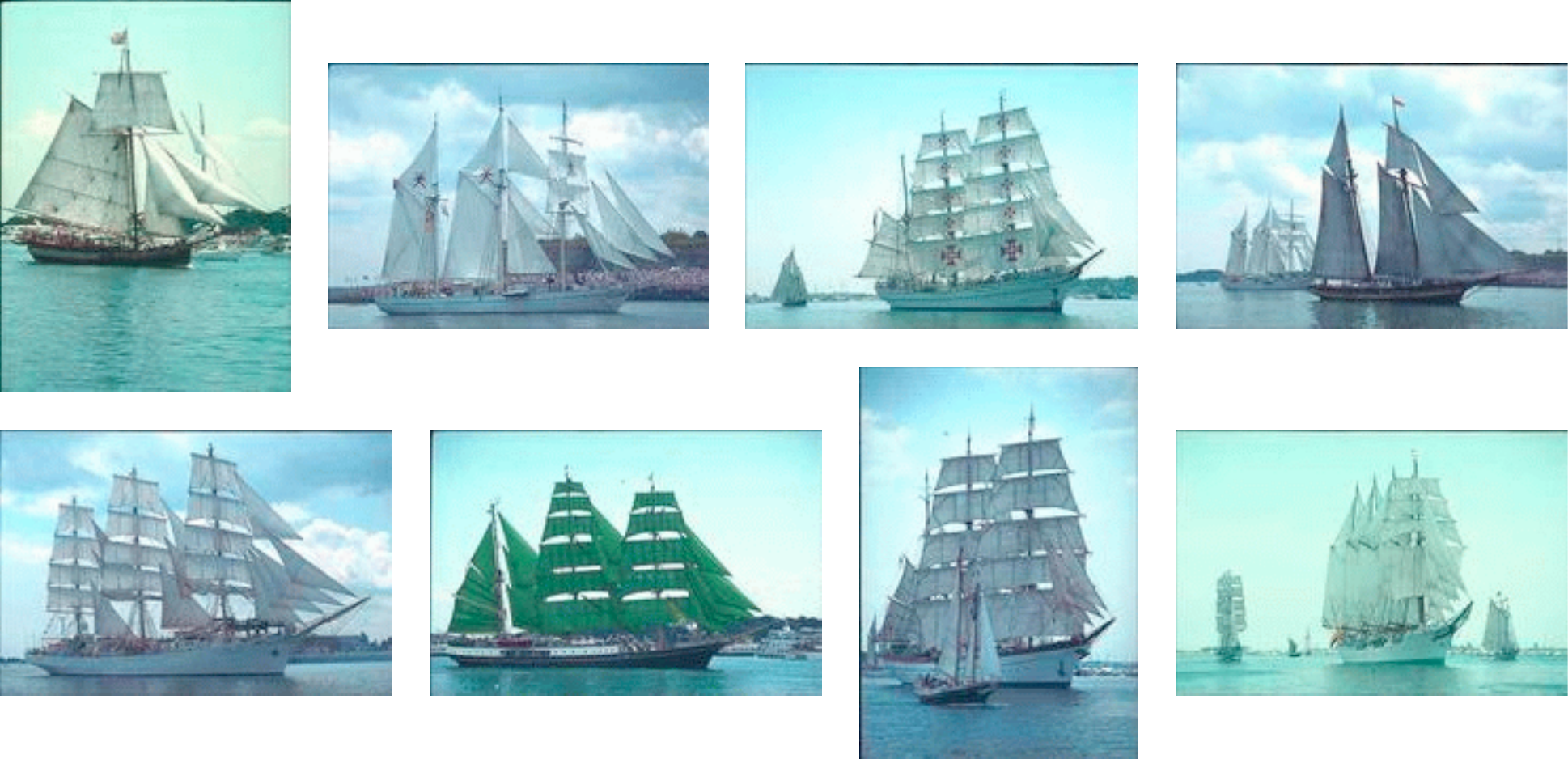}
\end{tabular}
\caption{A sample query (sub)image and its relevant answer set.}
\label{sailBoatAnswerset}
\end{figure*}

\begin{figure*}
\center
\begin{tabular}{c  c  c}
Rank: 1 & Rank: 12 & Rank: 18 \\
\includegraphics[width=2.5cm]{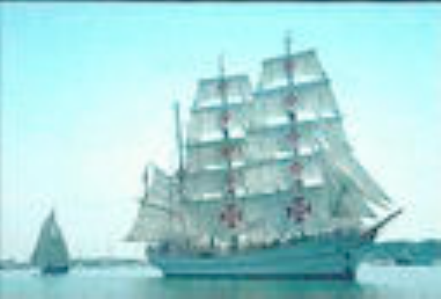} &
\includegraphics[width=2.5cm]{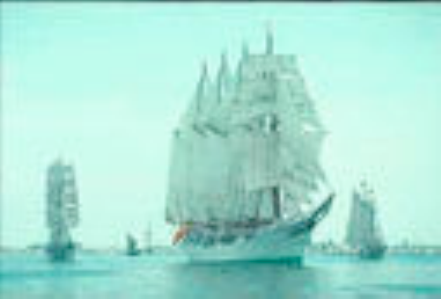} &
\includegraphics[width=2.5cm]{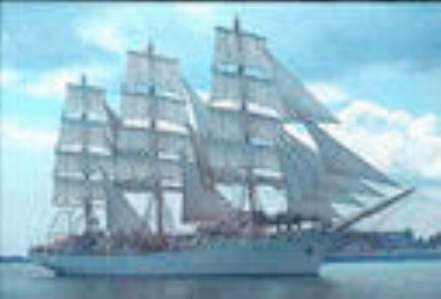} \\
\end{tabular}
\caption{Rank of the relevant images obtained after the first iteration
(no feedback given).}
\label{iteration1}
\end{figure*}

\begin{figure*}
\center
\begin{tabular}{c c c}
Rank: 1 & Rank: 2 & Rank: 3 \\
\includegraphics[width=2.5cm]{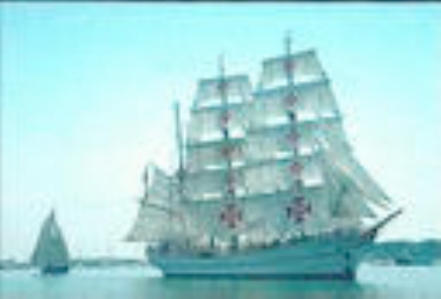} &
\includegraphics[width=2.5cm]{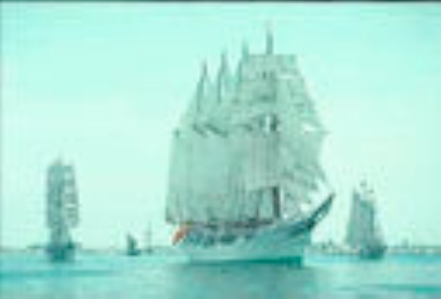} &
\includegraphics[width=2.5cm]{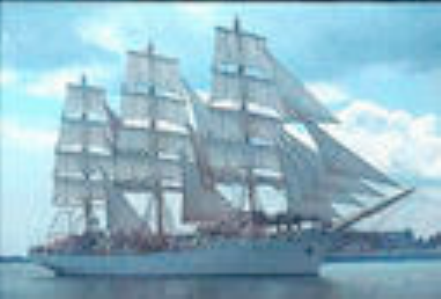} \\
Rank: 4 & Rank: 5 &  \\
\includegraphics[width=2.5cm]{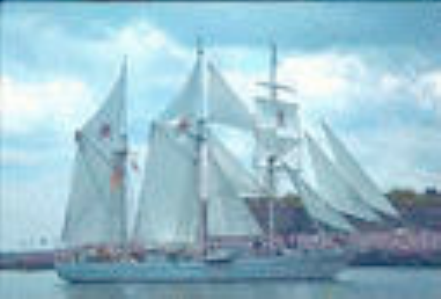} &
\includegraphics[width=2.5cm]{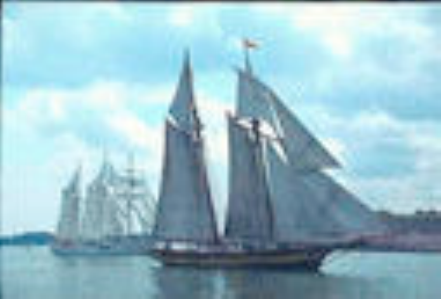} & \\
\end{tabular}
\caption{Rank of the relevant images obtained after the second iteration
(feedback given once).} 
\label{iteration2}
\end{figure*}

Most of the content-based image retrieval (CBIR) systems perform
retrieval based on a full image comparison, i.e., given a query image
the system returns overall similar images. This is not useful if users
are also interested in images from the database that {\em
contain} an image (perhaps an object) similar to a query
image. 
We call this searching process Content-Based 
sub-Image Retrieval (CBsIR), and it is defined as follows \cite{Sebe99}: 
given an image query
$Q$ and an image database $S$, retrieve from $S$ those images $Q'$ which
{\em contain} $Q$ according to some notion of similarity. 
To illustrate this consider Figure~\ref{sailBoatAnswerset} 
which displays an example query image and its relevant answer set.
Figures~\ref{iteration1} shows the images of such an answer set, 
and their respective ranks, retrieved within the top 20
matches after CBsIR is performed\footnote{
Incidentally, this is an example of an actual result obtained
using our prototype CBsIR system, available at
{\tt http://db.cs.ualberta.ca/mn/CBsIR.html}
}.   Note that the other
17 images returned are considered non-relevant to the query.
Now assume that the user is given the opportunity to mark those 3
images as relevant and all other 17 as irrelevant, i.e., the user
is allowed to provide {\em relevance feedback}.
Figure~\ref{iteration2} shows the relevant images retrieved (along
with their rank) after taking such feedback into account.
Note that all images previously obtained were ranked higher and
also new images were found and ranked high as well.

The sub-image
retrieval problem we consider is similar to region-based image retrieval
(RBIR), e.g. \cite{carson99,li00}, since the goal may also be to retrieve
images at object-level. However, there is a fundamental difference between 
these two.
The CBsIR problem is to search for an image, given as a whole, which is
contained within another image, whereas in RBIR one is searching for a
region, possibly the result of
some image segmentation. The former is more intuitive since users can
provide a query image as in traditional CBIR, and unlike the latter,
it
does not rely on any type of segmentation preprocessing.
Unfortunately,
automatic image segmentation algorithms usually lead to inaccurate
segmentation of the image when trying to achieve homogeneous visual
properties. Sometimes the obtained regions are only parts of a real
object and
should be combined with some neighbor regions so as to represent a
meaningful object. Thus, complex distance functions are generally used
to
compare segmented images at query time. Also, the number and
size of regions per image are variable and a precise representation of
the obtained regions may be storage-wise expensive.
Furthermore, since
region-based queries are usually performed after the image
segmentation and region description steps, it clearly puts some
restriction on the user's expression of his/her information need
depending on how good the segmentation results match the semantics of
images, even though the user can explicitly select any detected region as
query region. In those image retrieval systems where images are
heterogeneous, rich in texture, very irregular and variable in
contents, accurate regions are hard to obtain, making RBIR likely
to perform poorly. 

The main contribution of this paper is to realize CBsIR 
by employing relevance feedback, in order to capture the user's intentions
at query time.
As we discuss in the next section, relevance feedback is an interactive 
learning technique which has already been demonstrated to
boost performance in CBIR and RBIR systems.
Despite the great potential shown by relevance feedback, to the best of 
our knowledge there is no published research that uses it in the 
context of CBsIR, thus positioning our work as unique in this domain.

The remainder of this paper is organized as follows. In the next section
we discuss some related work.  We also
summarize the BIC method \cite{Stehling02} for CBIR and how we adopt it
for the CBsIR techniques system we propose. (As we shall discuss
BIC is used as a building block when modeling images within our proposed
approach.) Our retrieval strategy
uses query refinement as well as the incorporation of user's judgement,
via relevance feedback,
into the image similarity measure.  This forms the core contribution
of this paper and is detailed in Section~\ref{RF}. In
Section~\ref{RFtests} we present and discuss experimental
results, which support our claim of improved retrieval effectiveness.
Finally, Section~\ref{RFsummary} concludes the paper and offers
directions for future work.

\section{Related Work}

In this section we initially survey some of the work done 
using relevance feedback (also referred to as ``learning'') in
the context of CBIR and then in the context of RBIR.  
For the sake
of completeness we also briefly mention some research proposed
for single-pass, i.e., not considering relevance feedback, CBsIR.
Finally, we also
review the BIC technique for image representation 
since it will be used as an important building block
within our proposed approach.  

\subsection{Relevance Feedback within Traditional CBIR}

The key issue in relevance feedback is how to use positive and
negative examples to refine the query and/or to adjust the similarity
measure. Early relevance feedback schemes for CBIR were adopted from
feedback schemes developed for classical textual document retrieval.
These schemes fall into two categories: query point movement (query
refinement) and re-weighting (similarity measure refinement), both
based on the well-known vector model. 

The query point movement methods aim at improving the estimate of the
``ideal query point'' by moving it towards positive example points and
away from the negative example points in the query space. One frequently
used technique to iteratively update the query is the Rocchio's formula
\cite{Rocchio71}. It is used in the MARS system \cite{RHM97},
replacing the document vector by visual feature vectors. Another
approach is to update query space by selecting feature models. The
best way for effective retrieval is argued to be using a ``society''
of feature models determined by a learning scheme since each feature
model is supposed to represent one aspect of the image content more
accurately than others. 

Re-weighting methods enhance the importance of a feature's dimensions,
helping to retrieve relevant images while also reducing the importance of
the dimensions that hinder the process. This is achieved by updating
the weights of feature vectors in the distance metric. The
refinement of the re-weighting method in the MARS system is called the
standard deviation method.

Recent work has proposed more computationally robust methods that
perform global feature optimization. The MindReader retrieval system
\cite{ISF98} formulates a minimization problem on the parameter
estimating process. Using a distance function that is not necessarily
aligned with the coordinate axis, the MindReader system allows
correlations between attributes in addition for different weights on
each component. A further improvement over the MindReader approach
\cite{RH99} uses a unified framework to achieve the optimal query
estimation and weighting functions. By minimizing the total distances
of the positive examples from the revised query, the weighted average
and a whitening transform in the feature space are found to be the
optimal solutions. However, this algorithm does not use the negative
examples to update the query and image similarity measure; and
initially the user needs to input the critical data of training
vectors and the relevance matrix into the system. 

Tasks that can be improved as a result of
experience can be considered as a machine-learning task.  Therefore,
relevance feedback can be considered as a learning method --the system
learns from the examples provided as feedback by a user, i.e., his/her
experience, to refine the
retrieval results. The aforementioned query-movement method represented
by the Rocchio's formula and re-weighting method are both simple
learning methods. However, as users are usually reluctant to provide a
large number of feedback examples, i.e., the number of training samples
is very small.  Furthermore, the number of feature dimensions in CBIR systems is
also usually high. Thus, learning from small training samples in
a very high dimension feature space makes many learning methods, such
as decision tree learning and artificial neural networks, unsuitable
for CBIR. 

There are several key issues in addressing relevance feedback in CBIR
as a small sample learning problem. First, how to quickly learn from
small sets of feedback samples to improve the retrieval accuracy
effectively; second, how to accumulate the knowledge learned from the
feedback; and third, how to integrate low-level visual and high-level
semantic features in the query. Most of the research in literature has
focused on the first issue.  In that respect Bayesian
learning has been explored 
and has been shown advantageous compared with other
learning methods, e.g., \cite{VL99}. Active learning methods have been
used to actively select samples which maximize the information gain,
or minimize entropy/uncertainty in decision-making. These methods
enable fast convergence of the retrieval result which in turn
increases user satisfaction. Chen et al \cite{CLZ01} use Monte carlo
sampling to search for the set of samples that will minimize the {\em
expected} number of future iterations. Tong and Chang \cite{TC01} 
propose the use of SVM active learning algorithm to select the
sample which maximizes the reduction in the size of the version space
in which the class boundary lies. Without knowing apriori the class of
a candidate, the best search is to halve the search space each time.
In their work, the points near the SVM boundary are used to
approximate the most-informative points; and the most-positive images
are chosen as the ones farthest from the boundary on the positive side
in the feature space. 

\subsection{Relevance Feedback within RBIR}

Relevance feedback has been introduced in RBIR systems for a
performance improvement as it does for the image retrieval systems
using global representations. 

In \cite{JLZ03}, the authors introduce several learning algorithms 
using the adjusted 
global image representation to RBIR. First, the query point movement
technique is considered by assembling all the segmented regions of
positive examples together and resizing the regions to emphasize the
latest positive examples in order to form a composite image as the new query.
Second, the application of support vector machine (SVM) \cite{TC01} in
relevance feedback for RBIR is discussed. Both the one class SVM as a
class distribution estimator and two classes SVM as a classifier are
investigated. Third, a region re-weighting algorithm is proposed
corresponding to feature re-weighting. 
It assumes that important regions should appear more
times in the positive images and fewer times in all the images of the
database. For each region, measures of region frequency RF and inverse
image frequency IIF (analogous to the TF and IDF in text
retrieval \cite{BR99}) are introduced for the region importance. Thus the
region importance is defined as its region frequency RF weighted by the
inverse image frequency IIF, and normalized over all regions in an image.
Also, the feedback judgement is memorized for future use by calculating
the cumulate region importance. However, this algorithm only consider
positive examples while ignoring the effect of the negative examples in
each iteration of the retrieval results. Nevertheless, experimental 
results on a
general-purpose image database demonstrate the effectiveness of those
proposed learning methods in RBIR.

\subsection{CBsIR without Relevance Feedback}

The paper by Leung and Ng
\cite{Leung98} investigates the idea of either enlarging the query
sub-image to match the size of an image block obtained by the four-level
multiscale representation of the database images, or conversely
contracting the image blocks of the database images so that they become
as small as the query sub-image. The paper presents an analytical cost
model and focuses on avoiding I/O overhead during query processing time.
To find a good strategy to search multiple resolutions, four techniques
are investigated: 
the branch-and-bound algorithm, Pure Vertical (PV), Pure Horizontal (PH) 
and Horizontal-and-Vertical (HV). The HV strategy is argued to be the
best considering efficiency. However, the authors do not report clear
conclusions regarding the effectiveness (e.g., Precision
and/or Recall) of their approach. 

The authors of \cite{Sebe99} consider global feature extraction to
capture the spatial information within image regions. The average color
and the covariance matrix of the color channels in L*a*b color space are
used to represent the color distribution. They apply a three level
non-recursive hierarchical partition to achieve multiscale
representation of database images by overlapping regions within them.
Aiming at reducing the index size of these global features, a compact
abstraction for the global features of a region is introduced.
As well, a new distance measure between such abstractions is introduced 
for efficiently searching through the tiles from the multi-scale partition
strategy. This distance is called {\em inter hierarchical distance} (IHD) 
since it is taken between feature vectors of different hierarchical
levels of the image partition. The IHD index is a two dimensional vector
which consumes small storage space. The search strategy is a simple 
linear scan of the index file, which assesses the similarity between the
query image and a particular database image as well as all its
sub-regions using their IHD vectors. Finally, the minimum distance found
is used to rank this database image.  

In \cite{jie03} a new method called HTM (Hierarchical Tree
Matching) for the CBsIR problem was proposed. It has three main components: (1)
a tree structure that models a hierarchical partition of images into
tiles using color features,
(2) an index sequence to represent the tree structure (allowing fast access during the search
phase), and (3) a search strategy based on the tree
structures of both database images and the query image. 
Since the tree structure presented in \cite{jie03} is re-used 
in our work, we detail it in the following.

\begin{figure*}
\centering
\includegraphics[width=12cm]{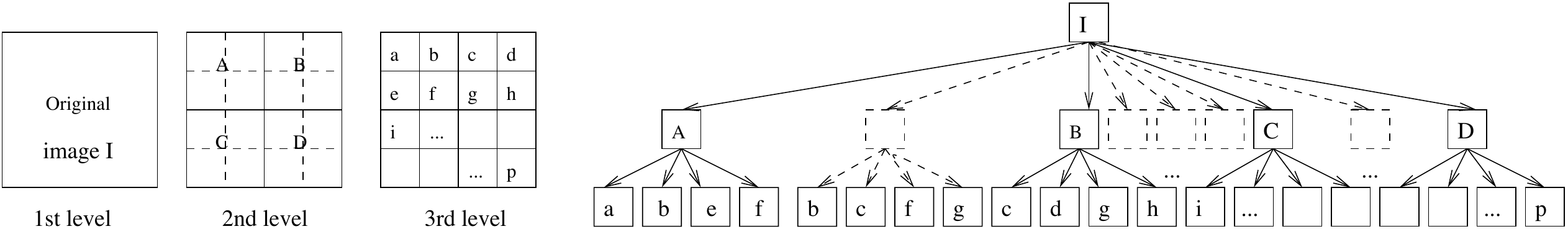}
\caption{Hierarchical partition of an image and the resulting tree
structure.}
\label{gridToTree}
\end{figure*}

To model an image, a grid is laid on it yielding a hierarchical
partition and tiles.
Although granularity could be arbitrary, we have obtained good results
using a 4$\times$4 grid
resulting in a three-level multiscale representation of the image
(similarly to what was  done in \cite{Leung98} and \cite{Sebe99}).
The hierarchial partition of an
image and its resulting tree structure are illustrated in
Figure~\ref{gridToTree}.
There are three levels in the hierarchical structure. The highest
level
is the image itself. For the second level the image is decomposed into
3$\times$3 rectangles with each side having half the length of the
whole
image, yielding 9 {\em overlapping} tiles. The lowest level consists
of
4$\times$9=36 rectangles, since each tile of the second level is
partitioned into 4 non-overlapping sub-tiles.
Note that, to exclude redundance in the CBsIR system,
only the indices of the 4$\times$4=16 unique tiles in the lowest
level are stored with a
small structure for relationship information. 
This tiling scheme is obviously not unique and as long as a well-formed
hierarchy of tiles is used to model the image the technique we
proposed can still be applied after corresponding adjustments.
The average color of the
image tiles in the RGB color space is associated to the nodes in the
tree stuctures for images\footnote{In this paper we do not use average
color, instead we use the BIC technique which provides an efficient and 
much more effective representation (c.f., Section~\ref{BIC}).}.
Thus, every database image is represented as
a series of tiles, each of which is mapped to a subtree of the tree
modeling the image.

An index sequence representing the predefined parent-child relationship (given by
the predefined order of sequence in the index) for the tree structure is stored
on secondary storage and used for fast retrieval. Details about the index
sequence structure can be found in elsewhere \cite{jie03}; in short, it resembles
a priority tree where the relative order among the tree nodes reflect the
relative order of the entries and which can be efficiently mapped onto an array
structure.  Such an structure allows one to efficiently traverse the necessary
indices for computing (sub)image similarity. The searching process is
accomplished by ``floating'' the tree structure of the query image over the full
tree structure of the candidate database image, shrinking the query's tree
structure so that it is comparable with the candidate database image's trees at
each level of the hierarchical structure. The minimum distance from tree
comparisons at all hierarchical levels, indicating the best matching tile from a
database image, is used as the distance between the database image and the query.
Differently from \cite{Sebe99}, the HTM search strategy considers local
information of images' tiles represented by leaf nodes in the subtree structures.
The average of distance values among the corresponding leaf nodes is taken for
the distance between the tree structures of query image and a certain tile of the
database image at any hierarchical level. Even though different datasets were
used, experiments detailed in \cite{jie03} strongly suggest that the proposed
approach yields better retrieval accuracy compared to \cite{Sebe99}, at the cost
of small storage overhead.

\subsection{The BIC-based Image Abstraction}
\label{BIC}

An straigthforward way to model images is to use its average color.
This is obviously not effective in any non-trivial situation.
Another simple, and in many situations cost-effective means is to 
use a global color histogram (GCH) (c.f., \cite{Lu99}).  A common critique 
to GCHs is that it is unable to capture any notion of spatial
distribution.  To address this several other approaches have been
proposed\footnote{A comprehensive survey thereof is beyond the scope of 
this paper.}, but they add complexity as a trade-off in order to gain
effectiveness.
Nevertheless, the use of color only, without any notion of 
spatial distribution, may be effective, if one is able to capture
other features of the images, e.g., texture.  That is exactly the
advantage of the BIC technique proposed in \cite{Stehling02} and which
we re-use within our proposal.

The image analysis algorithm of BIC classifies pixels 
as either border, when its color is the same as its neighboors, or
otherwise as interior, and two normalized histograms are computed 
considering only the border pixels and the interior pixels respectively.
That is, for each color two histogram bins exist: one in the border pixel
histogram and one in the interior pixel histogram.  This
allows a more informed color distribution abstraction and captures,
implicitly, a notion of texture.  

To illustrate the idea consider two
images, one composed of two equally sized solid color blocks 
of different colors, say C1 and C2, and another one where half of 
pixels of color have color C1 and are randomly distributed.  Likewise
the other half of pixels have color C2 and are also randomly
distributed. Clearly the BIC histograms of those
images are quite different, one will have almost only interior pixels
and the other will have almost only border pixels.  This will yield a 
low similarity measure, which is indeed the case.
Note that the global color histogram, a standard CBIR technique, for 
both images would be identical, misleading one to think the images
were very similar.
Note that the difference in the histogram suggests a very different
texture in the images, which, on top of the possible color
differences, enhances the capability of distinguishing among images
even further.

\begin{figure*} [htb!]
\center
\includegraphics[width=12cm]{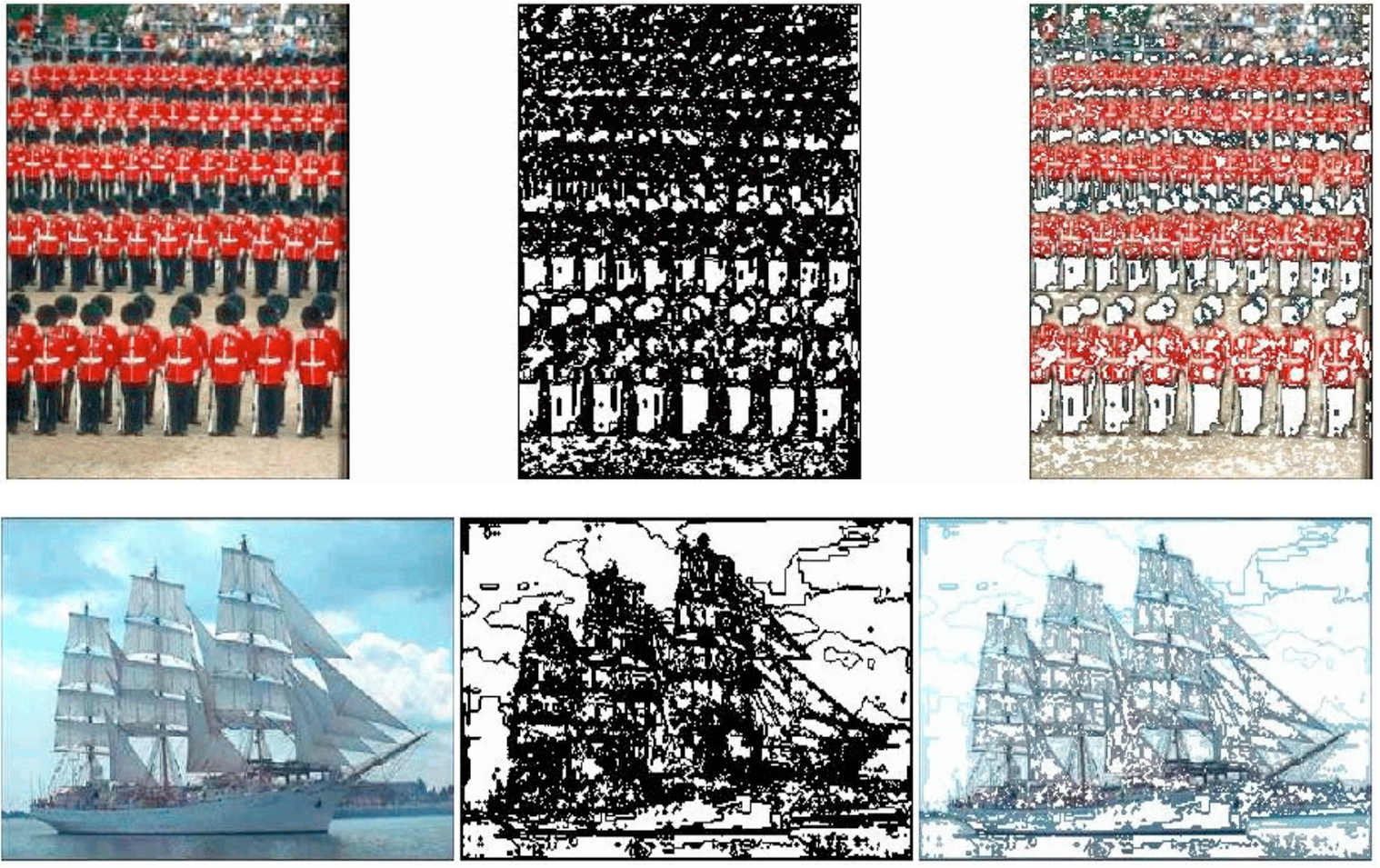}
\caption{Two examples of the result by the BIC pixel classification.}
\label{BICresults}
\end{figure*}

Figure~\ref{BICresults}\footnote{
C.f. {\tt http://db.cs.ualberta.ca/mn/BIC/bic-sample.html}
} shows two examples of images analyzed by border and interior pixels,
where the notion of capturing texture can be clearly seen.
The original images are at the left column. The resulting binary
images showing border pixels in black and interior pixels in white are
at the middle column. The images showing border pixels in the
corresponding original colors and interior pixel in white are at the
right column.

For histogram comparison within BIC, 
the $dLog$ distance function is used to diminish
the effect that a large value in a single histogram bin
dominates the distance between histograms, no matter the relative
importance of this single value \cite{Lu99,Nascimento02}. 
The basic motivation behind this is based on the observation that
classical techniques based on global color histograms treat all colors
equally, despite of their relative concentration. However, the
perception of stimulus, color in images in particular, is believed to
follow a ``sigmoidal'' curve \cite{Nascimento02}.  The more relative 
increment in a stimulus is perceived more clearly when the intensity
of the stimulus is smaller than when it is larger.  For instance,
a change from 10\% to 20\% of a color is perceived more clearly than
a change from 85\% to 95\%.
Indeed, it has been a well observed phenomena regarding many other
phenomena involving how sensitive one is (including animals) to
different stimuli \cite{falmagne86}.
Thus, the distance function is defined as: 
$dLog(a,b) =
\sum_{i=0}^{M}\vert f(a[i]) - f(b[i]) \vert$ where
$f(x) = \left \{ \begin {array}{ll}0 & \textrm{if $x=0$}\\ 1 & \textrm{if $0 < x \leq 1$}\\ \lceil log_{2}x \rceil +1 & \textrm{otherwise} \end{array} \right.$
\\
and $a[i]$ and $b[i]$ represent the $i^{th}$ bin of the
$M$ color histograms 
$a$ and $b$ respectively.  Note that if we normalize the histograms
bins in the [0, 255] range of integer values, instead of usual [0, 1] 
continuous
range, the $f(x)$ function will return integers
in the range [0, 9], requiring only $4$ bits of storage per histogram
bin.  This allows substantial reduction in storage, and yet
a reasonably fine discretization of the bins.

The BIC approach was shown in  \cite{Stehling02} to outperform
several other CBIR approaches and, as such, we adopt it in our
CBsIR proposal to extract and compare the visual feature
of each tile with the goal of improving the retrieval accuracy.

\section{Relevance Feedback for CBsIR}
\label{RF}

Despite the great potential of relevance feedback shown in
CBIR systems using global representations and in RBIR systems, to the
best of our knowledge there is no research that uses it within
CBsIR systems. In this section we present our solution for CBsIR
by using relevance feedback to learn the user's intention. Our
relevance feedback approach has three main components: (1) a tile
re-weighting scheme that assigns penalties to each tile of database
images and updates those tile penalties for all relevant images
retrieved at each iteration
using both the relevant (positive) and irrelevant (negative) images identified by the user; 
(2) a query refinement strategy that is based on the tile re-weighting
scheme to approach the most informative query according to the user's
intention; (3) an image similarity measure that refines the final
ranking of images using the user's feedback information. Each of these
components is explained in detail in the following subsections.

\subsection{Tile Re-Weighting Scheme}
Researches in RBIR \cite{JZL01,JLZ03} have proposed region
re-weighting schemes for relevance feedback.
In this research, we design our tile re-weighting scheme that 
specializes the technique
presented in \cite{JZL01} to accomodate our tile-oriented (not
region-oriented) HTM approach for CBsIR. It should be emphasized that
instead of considering all the images in the database to compute the
parameters for region weight \cite{JLZ03} (which is computationally
expensive), our tile re-weighting scheme uses only the positive and
negative examples identified by the user to update the {\em tile
penalty} of the positive images only, which is much more efficient.
Moreover, the region re-weighting scheme in \cite{JZL01} uses a
predefined similarity threshold to determine whether the region and
the image is similar or not, otherwise the comparison of region pairs
would become too expensive since images might consist of different and
large number of regions. This threshold is sensitive and subject to
change for different kinds of image datasets. Thus, how to obtain the
right threshold is yet another challenge for the relevance feedback
method in RBIR. However, our RF method for the CBsIR problem does not
need any threshold because the number of obtained tiles is the same
(and small) for each database image and there exists implicit
relationship between the tiles, which makes it easier to compare them.

In our system, the user provides feedback information by identifying
positive and negative examples from the retrieved images. The basic
assumption is that important tiles should appear more often in positive
images than unimportant tiles, e.g., ``background tiles'' should yield
to ``theme tiles'' in positive images. On the other hand, important
tiles should appear less often in negative images than unimportant
tiles. Following the principle of ``more similar means better matched
thus less penalty'', we assign a penalty to every tile that
represents the database image for the matching process. User's feedback
information is used to estimate the ``tile penalties'' for all positive
images, which also refines the final ranking of images. During
the feedback iterations, the user does not need to specify which tile of a
certain positive image is similar to the query, which would only make
the problem only simpler to solve at an additional cost to the user. 

Next, we introduce some definitions used to determine the tile
penalty and formalize the overall relevance feedback process. 

\smallbreak
\noindent
{\bf \em Definition 1}: The distance between two tiles $T_{a}$ and
$T_{b}$ from images $I_{a}$ and $I_{b}$ respectively, is: 
\begin{displaymath}
DT(T_{a},T_{b})= \\
\frac{\sum_{i=1}^{m} d(Feature(T_{a_{i}}),Feature(T_{b_{i}}))}{m}
\end{displaymath}
where $T_{a_{i}}$ and $T_{b_{i}}$ are sub-tiles 
of $T_{a}$ and $T_{b}$ respectively,
$m$ is the number of unique leaf nodes in the tiles' tree
structures at any hierarchical levels (if already at the leaf level,
$m=1$), the distance function $d$ is to be instantiated with some particular
measure based on the result of the feature extraction done by
the $Feature$ function on the tiles, e.g., BIC's {\em dLog()} function defined 
in the previous section. $\bullet$
\smallbreak
\noindent
{\bf \em Definition 2}: The penalty for a certain tile $i$ from a database image after $k$ iterations is defined as:
$TP_{i}(k), i=0, \cdots, NT$,
where $NT+1$ is the number of tiles per database image, and 
$TP_{i}(0)$ is initialized as $\frac{1}{NT+1}$. $\bullet$ 

For instance, in Figure~\ref{gridToTree}, $NT+1=1+9+16$, i.e.,
 is equal to the number of nodes in the tree structure representing the
hierarchical partition of a database image; for the lowest level, only
unique nodes count. 
\smallbreak
\noindent
{\bf \em Definition 3}: For each tile from a positive image, we define a measure of the distance $DTS$ between tile $T$ and an image set $IS=\{I_{1},I_{2},\cdots,I_{n}\}$. This reflects the extent to which the tile is consistent with {\em other} positive images in the feature space. Intuitively, the smaller this value, the more important this tile is in representing the user's intention.
\begin{displaymath}
DTS(T,IS)=
\left 
\{\begin{array}
{ll}
\sum_{i=1}^{n} exp(DT(T,I_{i}^{0})), \\ 
	\textrm{if $T$ is at full tree level}\\
\sum_{i=1}^{n} exp(min_{j=1..NT}DT(T,I_{i}^{j})),  \\
	\textrm{if $T$ is at the subtree level}
\end{array} 
\right.
\end{displaymath}
where $NT$ in this case is the number of tiles at the current subtree level.
$\bullet$

Assuming that $I$ is one of the identified positive example images, 
we can compute the tile penalty of 
image $I$ which consists of tiles $\{T_{0},T_{1},\cdots,T_{NT}\}$. The user provides positive and negative 
example images during each $k^{th}$ iteration of feedback, denoted respectively as $IS^{+}(k) = \{I_{1}^{+}(k),\cdots, I_{p}^{+}(k)\}$ 
and $IS^{-}(k) = \{I_{1}^{-}(k),\cdots,I_{q}^{-}(k)\}$, where $p+q$ is
typically much smaller than the size of the database.

Based on the above preparations, we now come to the definition of tile penalty.
\smallbreak
\noindent
{\bf \em Definition 4}: For all images (only being positive), the tile
penalty of $T_{i}$ after $k$ iterations of feedback is computed (and
normalized) as:
\begin{displaymath}
TP_{i}(k)=\frac{W_{i}\times DTS(T_{i}, IS^{+}(k))}{\sum_{j=0}^{NT}(W_{j} \times DTS(T_{j},IS^{+}(k))}
\end{displaymath}
where
$W_{i}=1-\frac{DTS(T_{i},IS^{-}(k))}{\sum_{j=0}^{NT}DTS(T_{j},IS^{-}(k))}$,
acts as a penalty, reflecting the influence of the negative examples.
$\bullet$

This implies the intuition that a tile from a positive example image
should be penalized if it is similar to negative examples. Basically, we compute the distances $DTS$ between a particular tile $T$ and the positive image set $IS^{+}$ as well as the negative image set $IS^{-}$ respectively to update the penalty of that tile from a positive example image. The inverse of the tile's distance from the negative image set is used to weight its corresponding distance from the positive image set. 

\begin{figure*} 
\centering
\includegraphics[width=8cm]{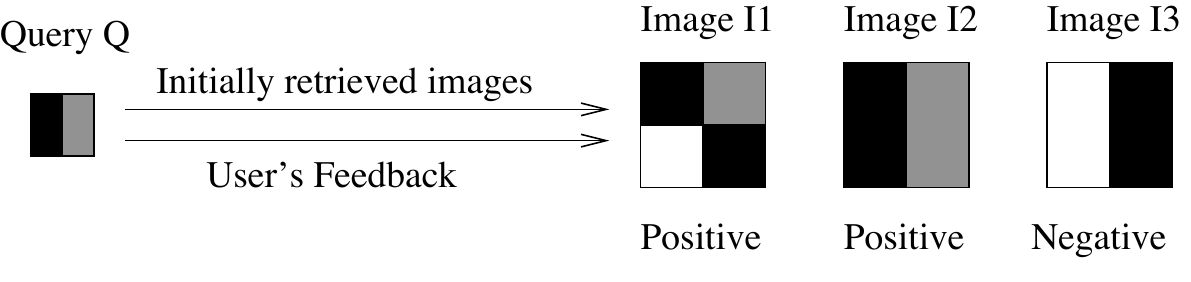}
\caption{Initial set of retrieved images with user's feedback.}
\label{initialRetrievalSet}
\end{figure*}

\begin{figure*} 
\centering
\includegraphics[width=12cm]{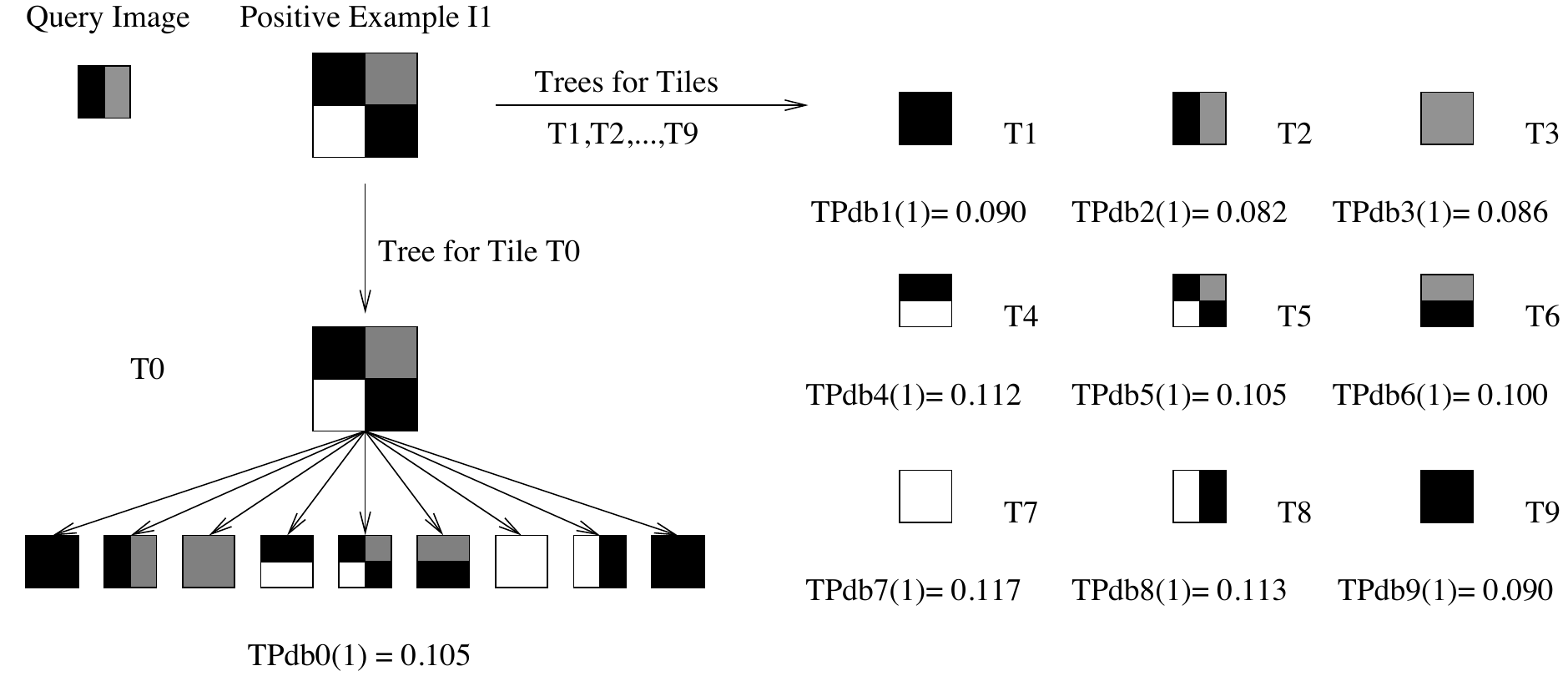}
\caption{Comparison of tile penalty for database image $I_{1}$ before and
after feedback.} 
\label{TPpos1}
\end{figure*}

Let us now illustrate the above methodology with a simple example, which also
motivates the notion of tile penalty. For simplicity, assume that the color
palette consists of only three colors: black, gray and white.
Figure~\ref{initialRetrievalSet} shows the top 3 retrieved images and the user's
feedback judgement. Image $I_{1}$ is marked as a positive example since it
actually contains the query image, which exactly represents the sub-image
retrieval problem we are dealing with. Image $I_{2}$ is also marked as a positive
example because it is the enlargement of the query image (and therefore
containing it as well). For the sake of illustration, assume a two-level
multi-cale representation of database images is used as in Figure~\ref{TPpos1}.

The tile penalties for tiles per 
database image are 
initialized as 0.1 for the 10 tiles, i.e.,
$TP_{i}(0)=0.1, i\in[0,9]$. Now, take tile $T_{1}$ for example. 
According to Definition 3, we need to compute the distances $DTS$
between $T_{1}$ and the positive/negative image set. In order to do
this, firstly, the distances between $T_{1}$ and all tiles at the
corresponding subtree levels of all the images in the positive/negative
image set should be obtained by Definition 1. Then, using
Definition 4 the new penalty of $T_{1}$ is updated from 0.1 to 0.090
correspondingly. The penalties for other tiles is updated in the same
way during each feedback iteration. We illustrate the new values 
of all tile penalties for database image $I_{1}$ as a positive example after
one feedback iteration in Figure~\ref{TPpos1}. We can see that after
the user provides feedback information, some tiles lose some
weight while others gain.
For instance, $T_{1}, T_{2}, T_{3}$ and $T_{9}$ receive less
penalties now because they only contain the color of grey and/or black
which is/are also in the query. $T_{0}, T_{4}, T_{5}, T_{7}$
and $T_{8}$ are penalized more since they all contain the color
white. The new weights for these tiles generally follow the trend that more 
percentage of white color more penalty. $T_{6}$, which is a rotation of the 
query image maintains its weight for this iteration. This means that our system
is to some extent also capable of perceiving changes such as rotation.
Besides, for a closer look at the updated tile penalties of positive image $I_{1}$,
$T_{1}$ receives more penalty than $T_{3}$ now although they are similar to the 
query image in the same degree. Note that, according to Definition 4, both 
the positive and the negative example images are used to calculate new tile penalties.
And we penalize a tile more if it is also somewhat more similar to the  
negative example images compared with other tiles in the positive example image. 
Thus it is reasonable that the tile penalty for $T_{1}$ appears higher than that for $T_{3}$
after feedback learning, since $T_{1}$ contains some black color which is also in the negative 
example image $I_{3}$ while $T_{3}$ contains only the grey color.

\subsection{Query Feature Update}

The relevance feedback process using query refinement strategy is based on the 
tile re-weighting scheme and all positive and negative example images. The main
concern is that we need to maintain as much as possible the original
feature of query image while introducing new feature elements that would
capture more new relevant images. Considering the hierarchical tree
structure of the query image, we use the most similar tile (with
minimum tile penalty) at every subtree level of each positive image to update the query feature at the corresponding subtree level.
\smallbreak
\noindent
{\bf \em Definition 5}: The updated query feature after $k$ iterations is:
\begin{displaymath}
qn_{l}^{k}[j]=\frac{\sum_{i=1}^{p}(1-TPmin_{i_{l}}(k))\times Pos_{i_{l}}^{k}[j]}{\sum_{i=1}^{p}(1-TPmin_{i_{l}}(k))}
\end{displaymath}
where $qn_{l}^{k}$ is the new feature with M dimensions for a subtree (tile)
at the $l^{th}$ level of the tree structure for the query image after
$k$ iterations,
$TPmin_{i_{l}}(k)$ is the minimum tile penalty for a subtree (tile)
found at the $l^{th}$ level of the tree structure for the $i^{th}$
positive image after $k$ iterations, $Pos_{i_{l}}^{k}$ is the feature
for the subtree (tile) with minimum tile penalty at the $l^{th}$ level
of the $i^{th}$ positive image's tree structure after $k$ iterations,
and $p$ is the number of positive images given by the user at this
iteration. $\bullet$

Intuitively, we use the weighted average to update the feature 
for a subtree (tile) of the query, based on the features of those tiles that 
have minimum tile penalties within respective positive images.  
In this way, we try to approach the optimal query that
carries the most information needed to retrieve as many relevant images
to the query as possible.

\subsection{Image Similarity}
\label{imageSimilarityRF}

With the updated query feature and tile penalties for positive images,
we can now define the distance between images and the query for ranking
evaluation at each feedback iteration. In order to locate the best match
to the query sub-image, our image similarity measure tries to find the
minimum from the distances between the database image tiles and the query (recall that both the database image and the query sub-image have been modeled by the tree structure in the same way) at corresponding hierarchical level in the tree structure, weighted by the tile penalty of corresponding database image tiles. 
\smallbreak
\noindent
{\bf \em Definition 6}: The distance between the (updated) query image
$Q$ and a database image $I$ at the $k^{th}$ iteration is:
\begin{displaymath}
DI_{k}(I,Q)=min_{i=0..NT}TP_{i}(k-1)\times DT(I_{i},Q_{j}) 
\end{displaymath}
where $NT+1$ is the number of all subtrees in the tree structure (tiles)
of a database image, and $TP_{i}(k-1)$ is the tile penalty for the $i^{th}$
tile of image $I$ after $k-1$ iterations. $\bullet$

For the comparison of full tree structures, $i=0$ and $j=0$,
indicating both the full tree structure of the database image and the query image. For the comparison of subtree structures, $i=1..N_{l}$ for each $1\leq j \leq (L-1)$,
where $N_{l}$ is the number of subtree structures at the $l^{th}$ level of the tree structure and $L$ is the number of levels of the tree structure, mapped from the hierarchical partition. $j$ indicates the subtree structure at a particular level of the query image's tree structure, as a result of shrinking the original query tree
structure to make the comparison with the subtree structures of database
images comparable. 

Finally, the overall relevance feedback process for the CBsIR system can be summarized in the following algorithm:
\begin{enumerate}
        \item 
	The user submits a query (sub)-image. 
	\item 
	The system retrieves the initial set of images using the
proposed similarity measure, which consists of database images
containing tiles similar to the query sub-image.
	\item 
	The system collects positive and negative feedback examples
identified by the user.
	\item 
	For each positive image, the tile penalties of those
tiles representing this image using positive examples and negative
examples is updated.
	\item 
	The system updates the query using positive images and their 
newly updated tile penalties.
	\item 
	The revised query and new tile penalties for database images
is used to compute the ranking score for each image and sort the
results.
	\item 
	Show the new retrieval results and, if the user wishes to
continue, go to step 3.
\end{enumerate}

\section{Experiments and Results}
\label{RFtests}

Before going further let us define the metrics we use to measure
retrieval effectiveness.
For certain applications, it is more useful that the system brings new
relevant images (found due to the update of query feature from
previous feedback) forward into the top range rather than keeping those
already retrieved relevant images again in the current iteration. For
other applications, however, the opposite situation applies, the user
is more interested in obtaining more relevant images during each
iteration keeping those s/he has already seen before. 
Given these observations, we use two
complementary measures for precision and recall as follows: 

\begin{enumerate}
  \item {\em New Recall}: the percentage of relevant images that were 
not in the set of the relevant images retrieved during previous
iterations over the number of relevant images in the answer set.
(Measured only after the first iteration, i.e., after the first feedback
cycle.) 
  \item {\em New Precision}: the percentage of relevant images that were 
not in the set of the relevant images retrieved during previous
iterations over the number of retrieved images at each iteration.
(Also measured after the first iteration.)
  \item {\em Actual Recall}: the percentage of relevant images at each iteration over the number of relevant images in the answer set. 
  \item {\em Actual Precision}: the percentage of relevant images at each iteration over the number of retrieved images at each iteration.
\end{enumerate}
The new recall
and precision explicitly measure the learning aptitude of the system;
ideally it retrieves more new relevant images as soon as
possible. 

Moreover, we also measure the total number of distinct relevant
images the system can find during all the feedback iterations. This is a
history-based measure that implicitly includes some relevant images
``lost'' (out of the currently presented images) in the process. We call 
them {\em cumulative recall} and {\em cumulative precision} defined as
follows:
\begin{enumerate}
  \item {\em Cumulative Recall}: the percentage of distinct relevant images from all iterations so far (not necessarily shown at the current iteration) over the number of relevant images in the predefined answer set. 
  \item {\em Cumulative Precision}: the percentage of distinct relevant images from all iterations so far over the number of retrieved images at each iteration.
\end{enumerate}

Table~\ref{measures} exemplifies the measures mentioned above, assuming
the answer set for a query contains 3 images A, B, C and the
number of returned (presented) images is 5. 
  
\begin{table*}
\caption{Cumulative/New/Actual Recall and Precision}
\center
\begin{tabular}{c|c|c|c|c}
\hline
Iteration& Relevant & Cumulative & New & Actual \\ 
& Retrieved & Recall/Precision & Recall/Precision &
Recall/Precision \\
\hline \hline
1& A & 33.33\%/20\% & --/-- & 33.33\%/20\%
\\ \hline
2& A & 33.33\%/20\% & 0\%/0\% & 33.33\%/20\% 
\\ \hline
3& B,C & 100\%/60\% & 66.67\%/40\% & 66.67\%/40\% 
\\ \hline
\end{tabular}
\label{measures}
\end{table*}

In addition to the above measures, we also 
evaluate storage overhead and query processing time.

We test the proposed relevance feedback approach 
using a heterogenous image dataset consisting of 10,150 color JPEG images: 
a mixture of the public Stanford10k\footnote{
{\tt http://www-db.stanford.edu/$\sim$wangz/image.vary.jpg.tar}.
} dataset and some images from one of COREL's CD-ROMs, each of which
falls into a particular category --we use 21 such categories\footnote{
The union of the images shown in {\tt
http://db.cs.ualberta.ca/mn/CBIRone/} and 
{\tt http://db.cs.ualberta.ca/mn/CBIRtwo/}
}. Some categories do not have rotated or translated images, but
others do.  On average, each answer set has 11 images, and none of the
answer sets has more than 20 images, which is the amount of images we
present to the user for feedback during each iteration.   It is
important to note that the queries
and answer sets are not part of the Stanford10k dataset in order to
minimize the probability that other images, not contained in the
expected answer set, could also be part of the answer but not
accounted for.
We manually crop part of a certain image from each of the above
categories to form a query image set of 21 queries (one for each
category). Images of the same categories serve as the answer sets for
queries (one sample query and its corresponding answer set are shown in 
Figure~\ref{sailBoatAnswerset}).
The size of the query image varies, being on average 18\% the size of the database images. The following performance results are collected from the online demo
available at {\tt http://db.cs.ualberta.ca/mn/CBsIR.html}. (An sample 
of the two initial iterations using our system is presented in the
Appendix.)

In our experiments, the maximum number of iterations explored is set to
10 (users will give feedback 9 times by pointing out which images are
relevant (positive)/irrelevant (negative) to the query) and we present
the top 20 retrieved images at each iteration. 
While within the same query session, the information collected 
at one step of the relevance feedback phase is used in the next step
(as indicated in the definitions presented in Section~3),
the information collected across different query
sessions is not integrated into
the search for the next queries --even if the very same query is
submitted to the system again.  I.e., we assume 
query sessions are independent; more specifically, once the user
goes to the initial page, all accumulated learning is cleared.
This consideration is based on the
observation of the subjectivity of human perception and the fact that
even the same person could perceive the same retrieval result
differently at different times.

As discussed earlier we use BIC histograms to model the contents of
an image tile.  The number of quantized colors in such histograms
is therefore a parameter for BIC.  We use two different values for this
parameter, 16 and 64 colors, in order to evaluate the influence of
the underlying tile model on the overall retrieval effectiveness.

Table~\ref{originalIterBIC} shows how many, on average,
iterations were necessary to have the original image (the one from
which the query sub-image was extracted) placed within the top 20 images.
It is clear that using 64 quantized colors is more efficient,
as the hit rate of the original images is almost optimal. 
Even though this trend, i.e., the more colors the better the
retrieval, is fairly intuitive, it is interesting to 
see that this advantage does not grow linearly with the number
of colors across all experiments.  That is to say, that even using
a low number of colors one can still obtain fairly good results.

\begin{table*}
\caption{How fast the original is ranked within the top 20 images.}
\centering
\begin{tabular}{c|c}
\hline
\# of colors in BIC & Average \# of iterations needed 
\\ \hline \hline
64 quantized colors & 1.1 
\\ \hline
16 quantized colors & $>$2.3  
\\ \hline
\end{tabular}
\label{originalIterBIC}
\end{table*}

\begin{figure} 
\center
\includegraphics[width=8cm]{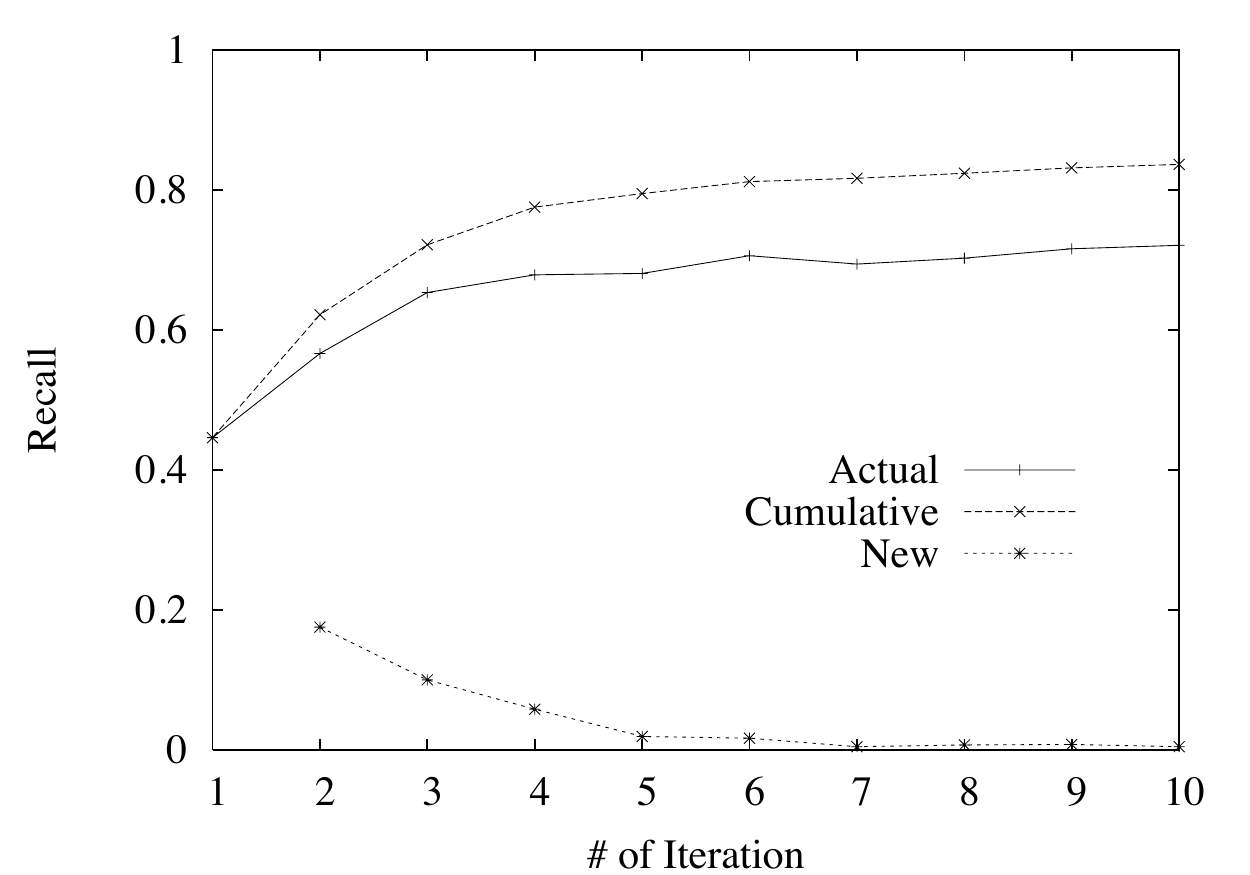}
\caption{Effectiveness measures by actual recall, cumulative recall and new recall using 64 quantized colors in BIC.}
\label{recall64colorsBIC}
\end{figure}

\begin{figure} 
\center
\includegraphics[width=8cm]{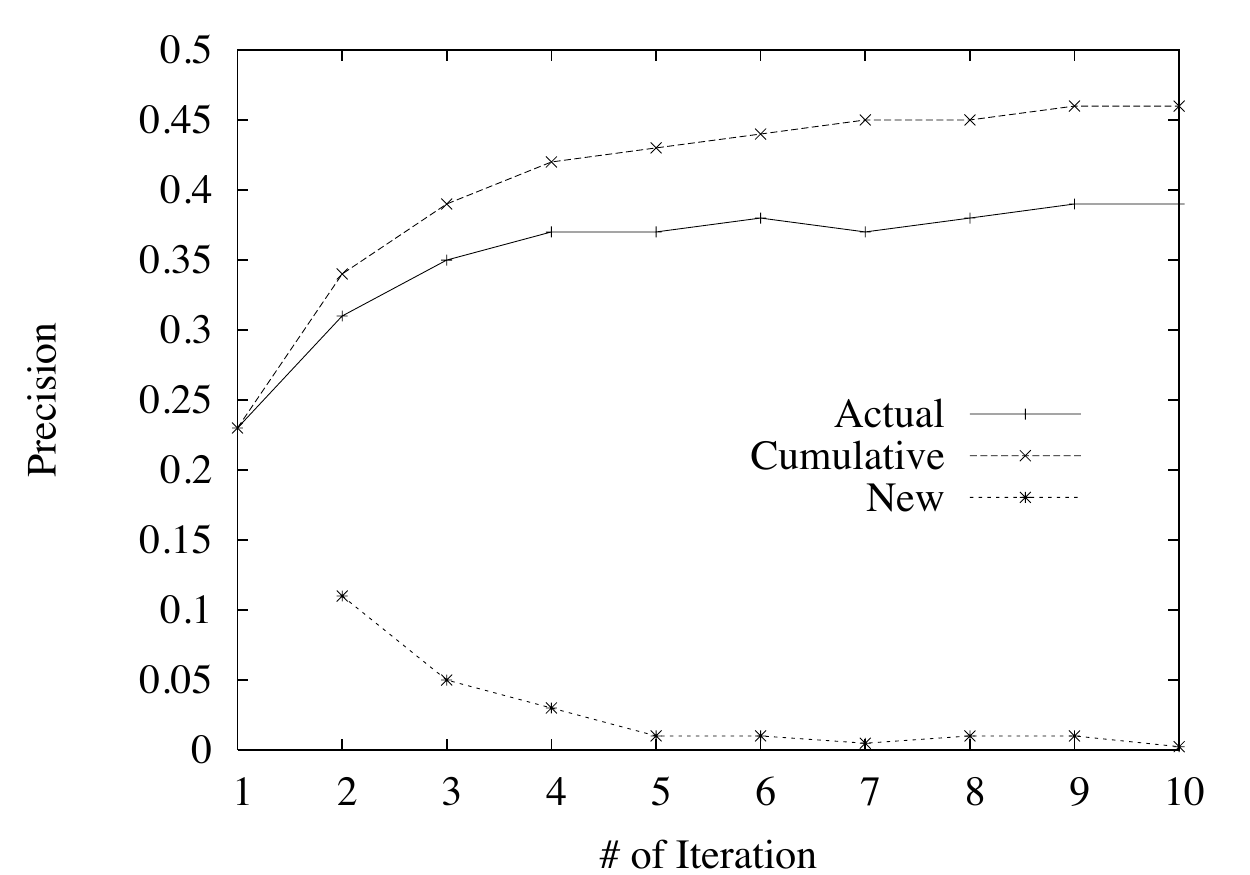}
\caption{Effectiveness measures by actual precision, cumulative precision and new precision using 64 quantized colors in BIC.}
\label{precision64colorsBIC}
\end{figure}

The retrieval accuracy using 64 quantized colors
is shown in
Figure~\ref{recall64colorsBIC} and Figure~\ref{precision64colorsBIC}.
As it can be clearly seen, after 5 iterations the system has already 
learned most of the information it could learn, i.e., the information
gain (given by the new recall and new precision curves) is nearly null.
On the other hand, after only 5 iterations the actual recall and actual
precision values increased by 55\% and 60\% respectively.
It is also noteworthy to mention that the stable actual precision value
of nearly 40\% is not as low as it may seem at first. 
The answer sets have
an average of 11 images and since the user is presented with 20
images,
the maximum precision one could get (on average) would be about
50\% as almost half of the displayed images could not be considered
relevant by construction.
This interpretation leads to the proposal of the following measure:

\begin{itemize}
\item
{\em Normalized Precision}: the actual precision over the maximum
possible actual precision value.
\end {itemize}

Interestingly enough, careful consideration of such a measure 
shows that is equivalent to the usual notion of (actual) recall.
Indeed,
consider $R$ and $A$ to
be the sets of relevant answers and the retrieved answers with respect
to a given query.  The actual precision is then defined as
$|R \cap A|/|A|$.  The maximum precision value one can obtain is
$|R|/|A|$.  When the former is divided by the latter one obtains
$|R \cap A|/|R|$ which is precisely the definition of actual
recall.

This leads to the argument that precision-based
measures are not well suited for this type of scenario, where
non-relevant images are very likely to be included in the answer
set regardless of their relevance.  The actual recall,
being concerned only with the relevant images is a more
realistic measure.
Under this argument, 70\% of stable actual recall (or
normalized precision) after 5 iterations seems quite reasonable.

We also
obtained about 85\% for cumulative recall and about 50\% for cumulative
precision. The reason for the higher values than those for actual 
recall and actual precision is because some relevant images that may be 
``lost'' in subsequent iterations are always accounted for in these measures.

\begin{figure} 
\center
\includegraphics[width=8cm]{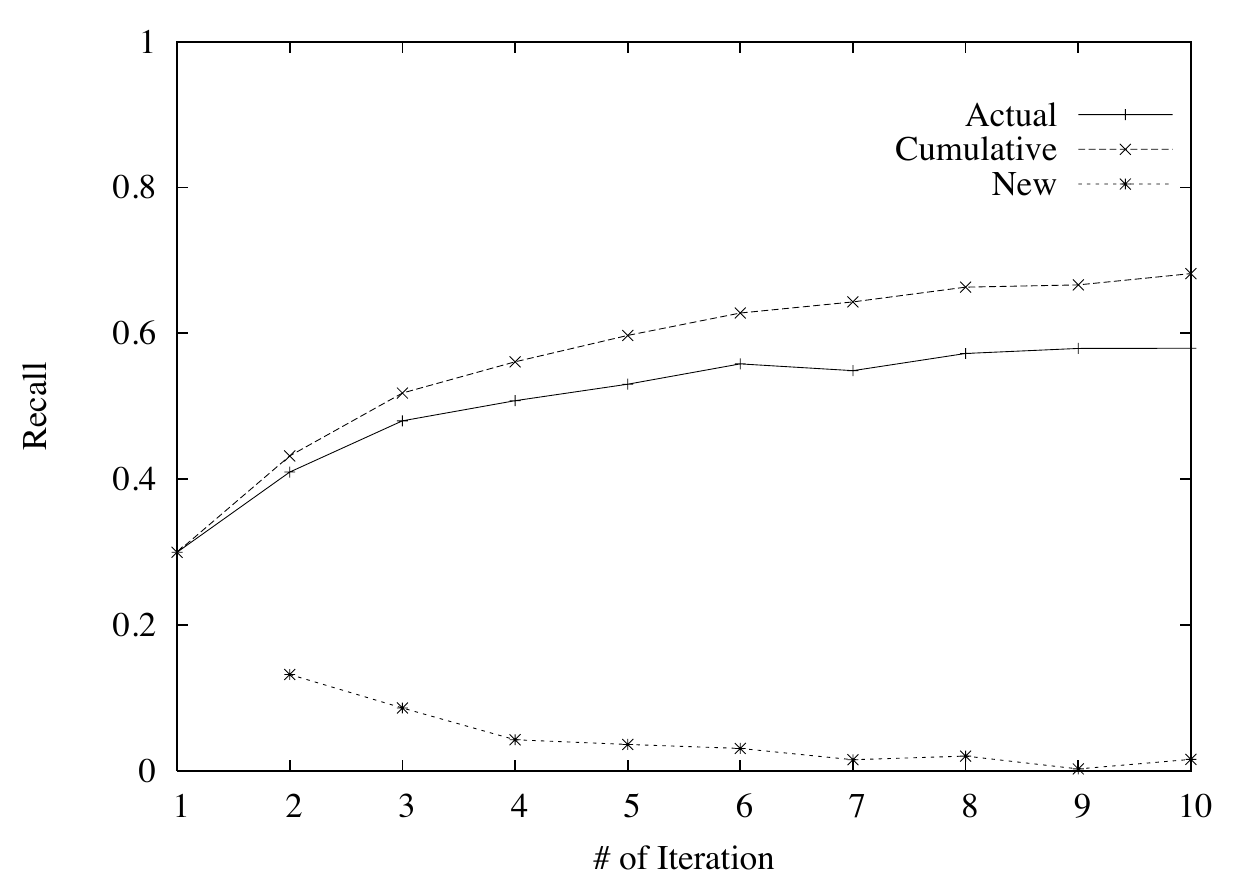}
\caption{Effectiveness measures by actual recall, cumulative recall and new recall using 16 quantized colors in BIC.}
\label{recall16colorsBIC}
\end{figure}

\begin{figure} 
\center
\includegraphics[width=8cm]{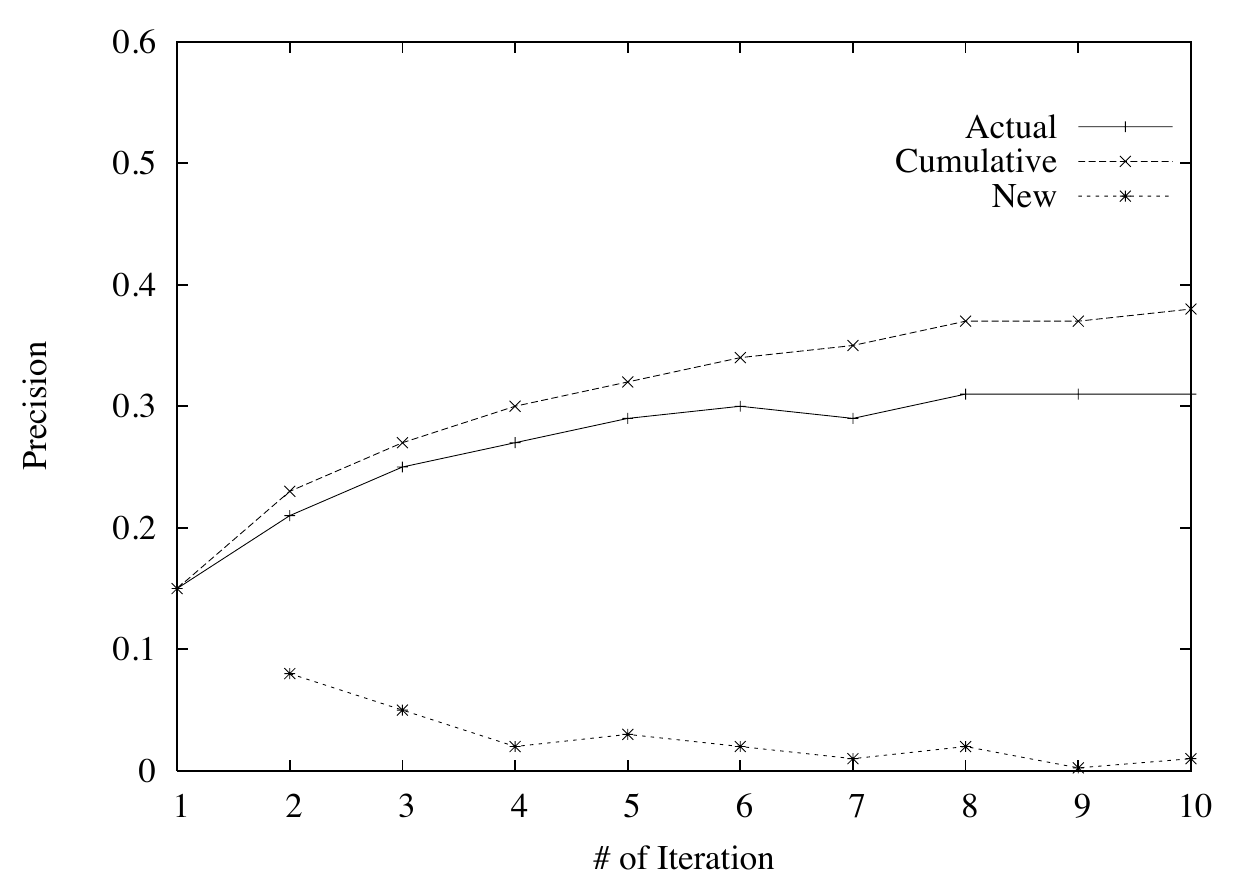}
\caption{Effectiveness measures by actual precision, cumulative precision and new precision using 16 quantized colors in BIC.}
\label{precision16colorsBIC}
\end{figure}

Using 16 quantized colors, as one would expect,
yields less accuracy than using 64 quantized colors.
However, an interesting aspect shown in Figures~\ref{recall16colorsBIC} and
\ref{precision16colorsBIC} is that even though, the amount of
information (i.e., number of colors) was reduced by 75\%, the
effectiveness was reduced by at most 10\% compared to the values in 
Figures~\ref{recall64colorsBIC} and \ref{precision64colorsBIC}.
The cost of the loss of information is more clear when looking 
at the ``learning aptitude.''   Using 16 colors required twice as 
many iterations in order to bring the curves to a stable state.
Still, this show a sublinear dependence on the number of colors:
using 4 times more colors yields only 10\% more effectiveness
and 2 times faster learning.

\begin{figure} 
\center
\includegraphics[width=8cm]{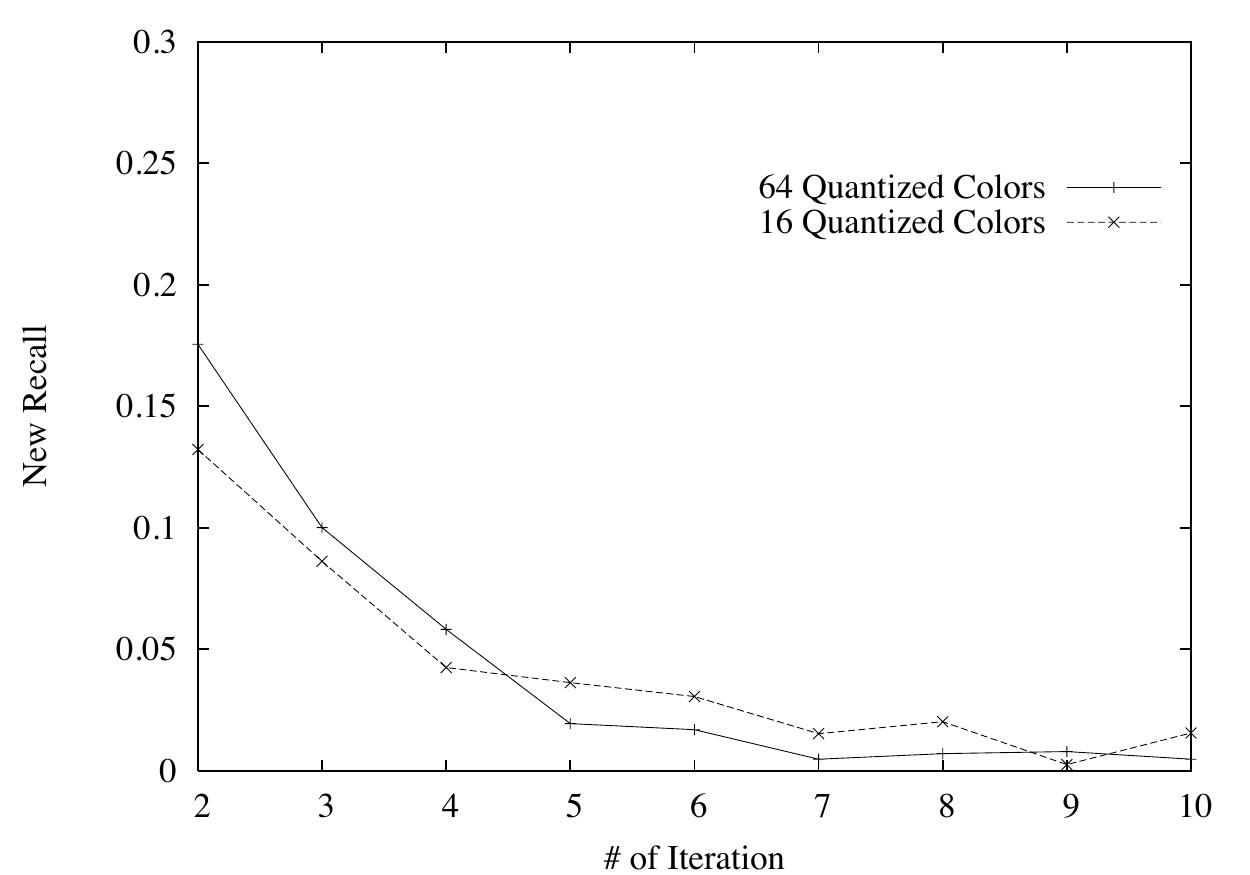}
\caption{New recall (defined from the second iteration) comparison using 64 quantized colors and 16 quantized colors in BIC.}
\label{newRecallCompare}
\end{figure}

\begin{figure} 
\center
\includegraphics[width=8cm]{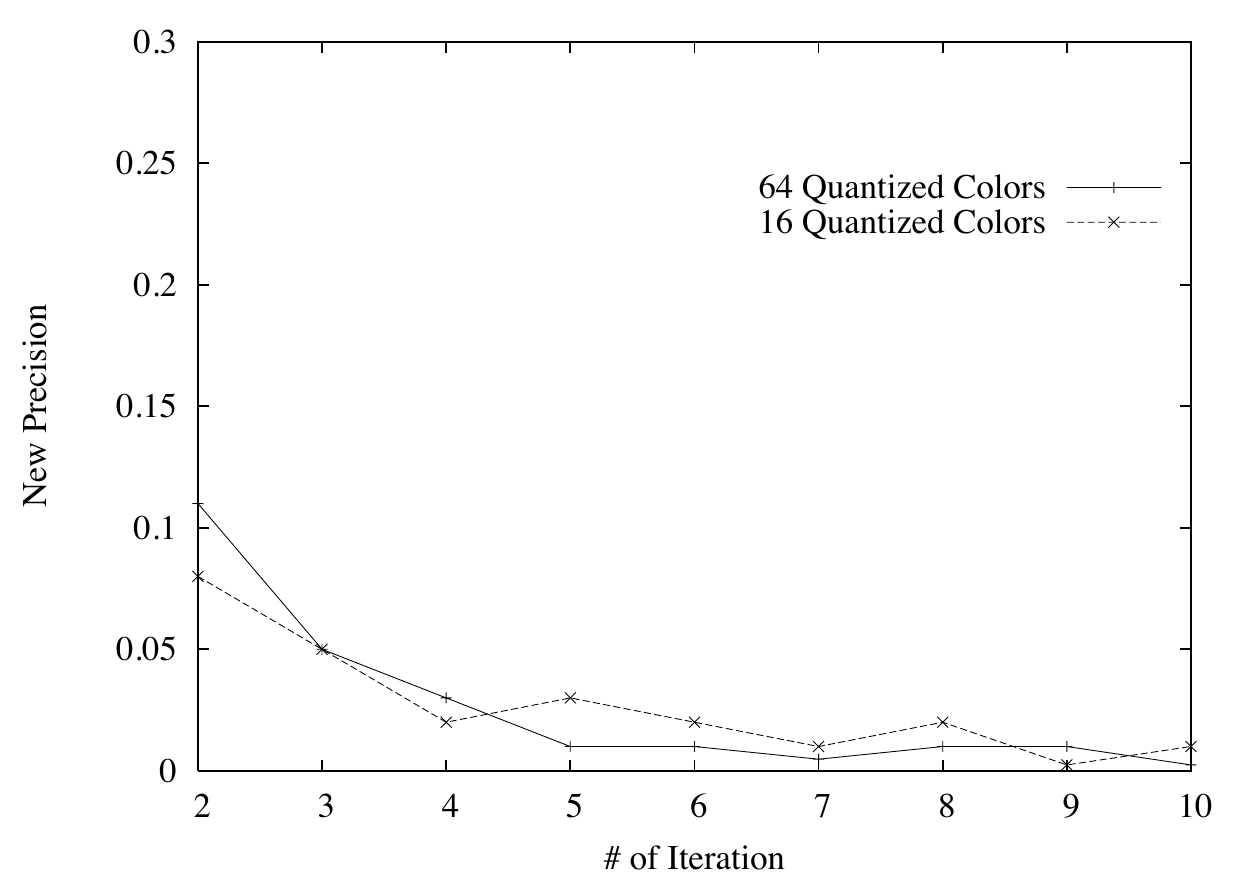}
\caption{New precision (defined from the second iteration) comparison using 64 quantized colors and 16 quantized colors in BIC.}
\label{newPrecisionCompare}
\end{figure}

Another interesting observation, which supports the main advantage 
of using more color for tile abstraction,
can be seen when comparing the new precision and recall curves
using different numbers of colors directly 
(Figures~\ref{newRecallCompare} and \ref{newPrecisionCompare}).
Up until the 4th or 5th iteration usign 64 colors yields
higher values, meaning that it is learning faster, after that
point, it has learned basically what it could have learn.
On the other hand the curve for using 16 colors shows that the
method is still learning.

\begin{figure} 
\center
\includegraphics[width=8cm]{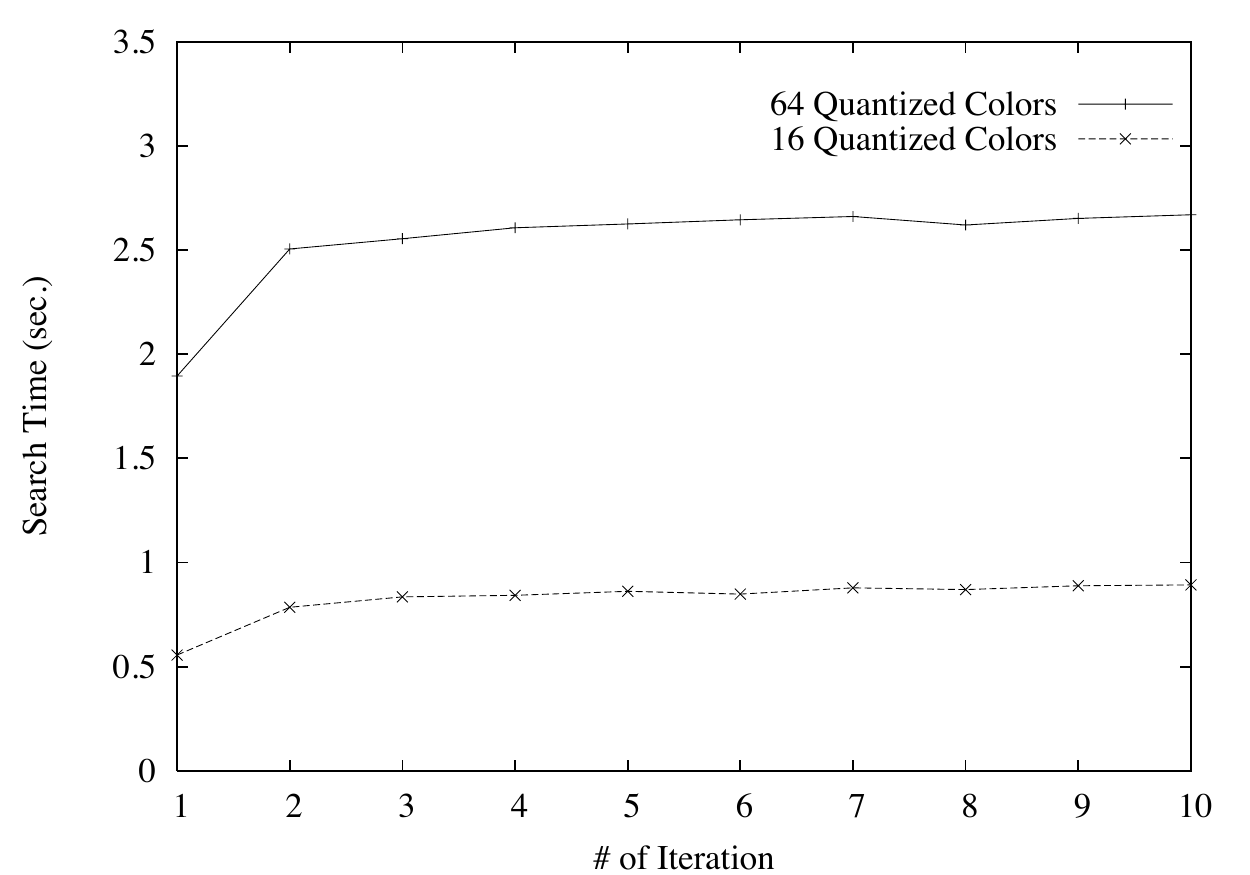}
\caption{Comparing query processing efficiency using different BIC histograms at each iteration.}
\label{searchTime}
\end{figure}

Figure~\ref{searchTime} shows the average time required
to process a query during each iteration, i.e., to access all
disk-resident data, complete the learning from the user's feedback at
the current iteration (not applicable to the first iteration), obtain
the distance between the query image and database images and sort them
by their resulting ranks. The first iteration takes, on average,
slightly less than 2 seconds when using 64 quantized colors and 0.6
second when using 16 quantized colors, whereas each subsequent
iteration requires about 2.5 seconds and 1 second respectively for the
two feature representations.  
This slight increase is due to the overhead for computing and updating
the tile penalties at each iteration. 
As well, note that the gain in speed is proportional to the smaller
number of colors used, i.e., 
using 64 colors yields a performance about times slower than
using only 16 quantized colors. 

Extracting image features from the image database, applying the
BIC method, and generating the metadata file
requires about 0.15 secs/image on a
computer running Linux 2.4.20 with AMD Athlon XP 1900+ CPU and 1GB of
main memory and is independent of the number of colors used --this 
procedure can be done off-line and should not be considered part of
query processing overhead. 

Finally, the
storage cost for the disk-resident metadata is 10.5 MB (only about
20\% the size of the image database), while using 16 quantized colors
needs proportionally less storage, namely 2.7 MB, again proportional
to the representation overhead.

\section{Conclusions}
\label{RFsummary}

In this paper we have shown, for the first time, how relevance feedback can 
be used to improve the performance of CBsIR. We presented 
a relevance feedback-based technique, which is
based on a tile re-weighting scheme that assigns penalties to each
tile of database images and updates those of all relevant images using
both the positive and negative examples identified by the user.
The user's feedback is used to refine the image
similarity measure by weighting the tile distances between the query
and the database image tiles with their corresponding tile
penalties. We combine this learning method with the BIC approach for
image modeling to improve the performance of content-based sub-image
retrieval. Our results on an image database of over 10,000 images suggest 
that the learning method is quite effective for CBsIR.  While using 
less colors within BIC reduce storage overhead and improve speedup query
processing it does not affect substantially retrieval efficiency in
the long term.  The main drawback is the system take longer to ``learn''
making the overall retrieval task a longer one.

A few possible venues for further investigation include the design of disk based
access structure for the hierarchical tree (to enhance the scalability for larger
databases), the use of better (more powerful yet compact) representation for the
tile features, possibly removing the background of the images and the
incorporation of more sophisticated machine learning techniques to shorten the
gap between low-level image features and high-level semantic contents of images
so as to better understand the user's intention.

\section*{Acknowledgments} A preliminary version of this paper appeared in the
Proceedings of the 2nd ACM Intl. Workshop on Multimedia Databases. This work was
supported in part by the Natural Sciences and Engineering Research Council
(NSERC), Canada, the Alberta Informatics Circle of Research Excellence (iCORE),
and the Canadian Cultural Content Management Research Network, a Network financed
through Heritage Canada's New Media Research Networks Fund.


\section*{Appendix}

\begin{figure} [htb]
\center
\includegraphics[width=12cm]{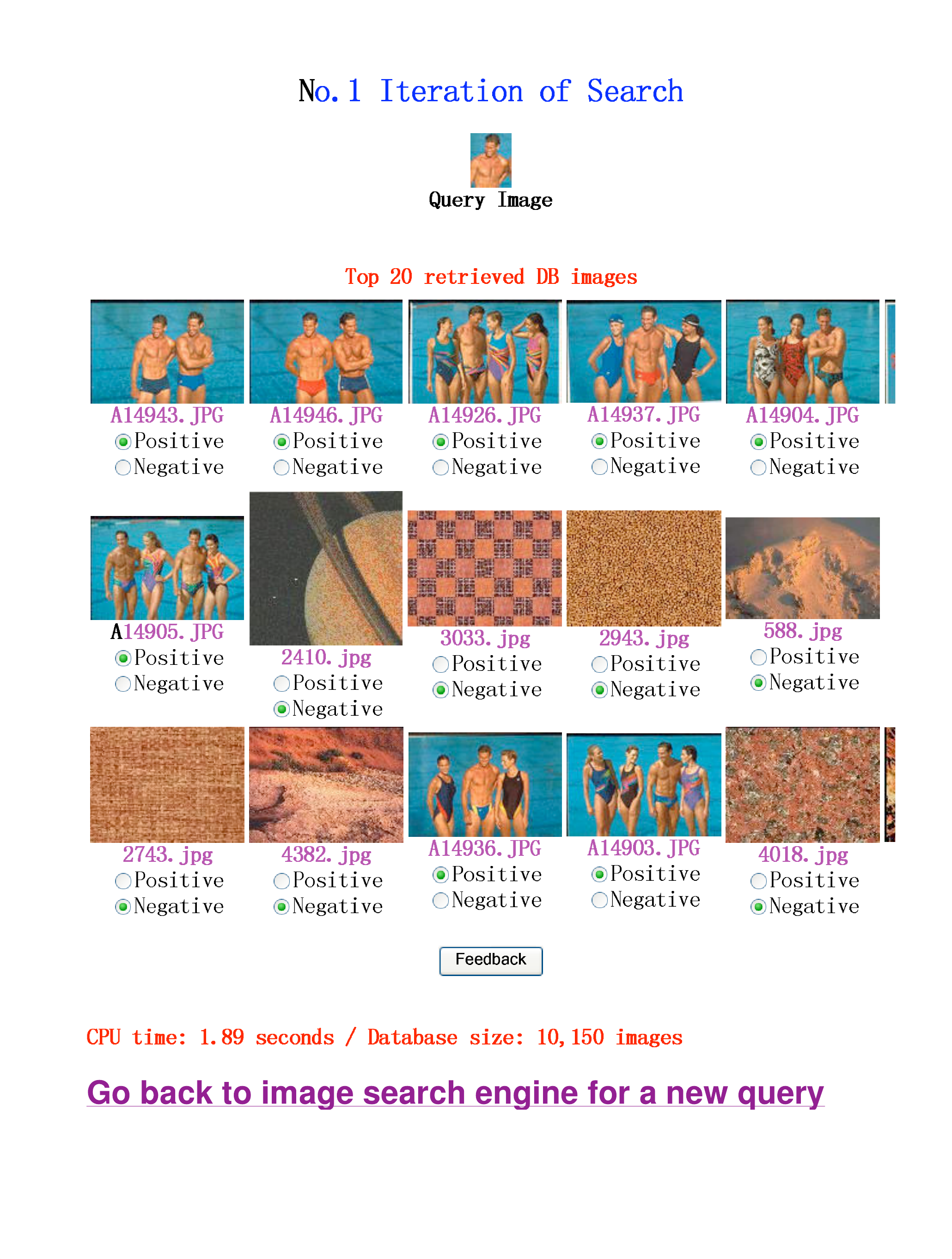}
\caption{Online demo using 64 colors for a
sample query after the first iteration (no feedback considered)}.
\label{swimmerIter1_64}
\end{figure}

\begin{figure} [htb]
\center
\includegraphics[width=12cm]{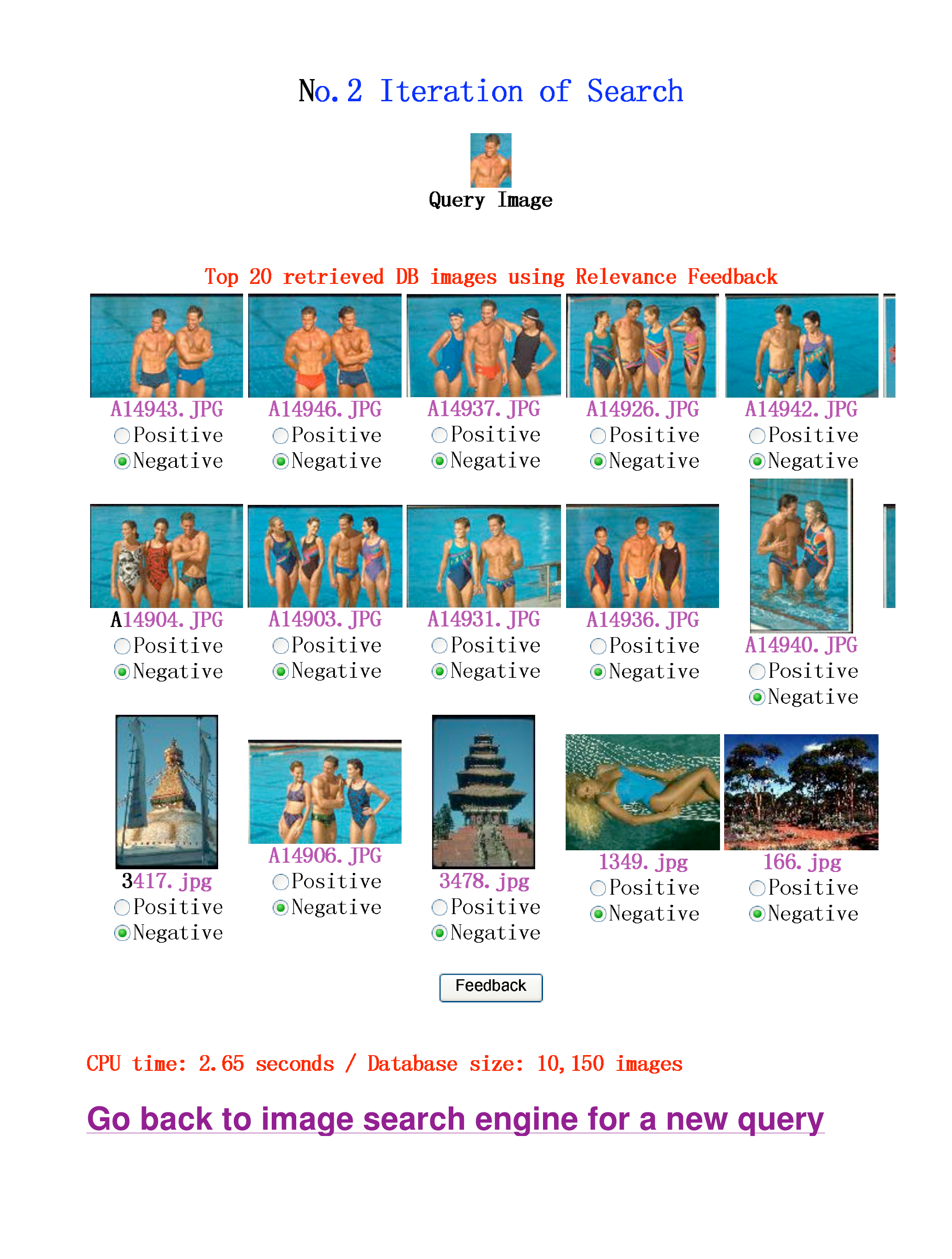}
\caption{Results after second iteration, i.e., feedback provided
once.}
\label{swimmerIter2_64}
\end{figure}

Figures~\ref{swimmerIter1_64} and \ref{swimmerIter2_64} offer a
of screenshots of the online demo for a sample query during the 
first two iterations.

\end{document}